\RequirePackage[2020-02-02]{latexrelease}
\documentclass[preprint,aps,tightenlines,showkeys,nofootinbib,superscriptaddress]{revtex4}
\usepackage{latexsym,amssymb,amstext,amsmath,bm}
\usepackage[dvips]{graphicx}
\usepackage[dvips]{color}
\newcommand{\beq}{\begin{eqnarray}}
\newcommand{\eeq}{\end{eqnarray}}
\newcommand{\bea}{\begin{eqnarray*}}
\newcommand{\eea}{\end{eqnarray*}}
\newcommand{\eq}{eqnarray}

\newcommand{\al}{{\alpha}}
\newcommand{\be}{{\beta}}
\newcommand{\ci}{\cite}
\newcommand{\ga}{{\gamma}}
\newcommand{\Ga}{{\Gamma}}
\newcommand{\ep}{{\epsilon}}

\newcommand{\de}{{\delta}}

\newcommand{\la}{{\lambda}}
\newcommand{\La}{{\Lambda}}

\newcommand{\si}{{\sigma}}
\newcommand{\Si}{{\Sigma}}
\newcommand{\ka}{{\kappa}}
\newcommand{\om}{{\omega}}
\newcommand{\Om}{{\Omega}}
\newcommand{\pa}{{\partial}}
\newcommand{\no}{{\nonumber}}
\newcommand{\f}{\frac}
\newcommand{\ra}{\rightarrow}

\begin{document}

\preprint{arXiv:2112.00576v4 [hep-th]}
\title{
Symmetries and Conservation Laws in Ho\v{r}ava 
Gravity}

\author{Deniz O. Devecio\u{g}lu \footnote{E-mail address: dodeve@gmail.com}}
\affiliation{School of Physics, Huazhong University of Science and Technology,
Wuhan, Hubei,  430074, China }

\author{Mu-In Park \footnote{E-mail address: muinpark@gmail.com, Corresponding author}}
\affiliation{ Center for Quantum Spacetime, Sogang University,
Seoul, 121-742, Korea }
\date{\today}

\begin{abstract}
{Ho\v{r}ava gravity has been proposed as a renormalizable quantum gravity without the
ghost problem through anisotropic scaling dimensions which break Lorentz symmetry in UV.
In the Hamiltonian formalism, due to the Lorentz-violating terms, the constraint structure
looks quite different from that of general relativity (GR) but we have recently found that ``{\it there exists
the case where we can recover the same number of degrees of freedom as in
GR}", in a rather general set-up.
In this paper, we study its Lagrangian perspectives and examine the full
diffeomorphism ({\it Diff}) symmetry and its associated conservation laws
in Ho\v{r}ava gravity. Surprisingly, we find that the {\it full} {\it Diff} symmetry in the action can also be
recovered when a certain condition, called {``super-condition"},
which {\it super-selects} the Lorentz-symmetric sector in Ho\v{r}ava gravity,
is satisfied. This indicates that the broken Lorentz symmetry, known as ``foliation-preserving" {\it Diff}, is just an {\it apparent} symmetry of the Ho\v{r}ava gravity action and rather its ``{\it full action symmetry can be as large as the {\it Diff} in GR} ". The super-condition exactly corresponds to the {\it tertiary}
constraint in Hamiltonian formalism which is the second-class constraint and
provides a non-trivial realization of the Lorentz symmetry otherwise
being absent apparently.
From the recovered Lorentz symmetry in the action,
we obtain the
conservation laws with the Noether currents as in covariant theories. The general formula for the conserved Noether charges reproduces the mass of four-dimensional static black holes with an {\it arbitrary} cosmological constant in Ho\v{r}ava gravity, and is independent of ambiguities associated with the choice of asymptotic boundaries. We also discuss several challenging problems,
including its implications to Hamiltonian formalism, black hole
thermodynamics, radiations from colliding black holes.}
\end{abstract}


\maketitle

\newpage

\section{Introduction}
The
renormalizable
gravity without the ghost problem has been studied
by considering
a 
gravity action {\it \`{a} la} Ho\v{r}ava, Lifshitz, and DeWitt (HLD) \ci{Lifs,DeWi,Hora:2009} in $(D+1)$ dimensions (up to some boundary terms),
\begin{align}
S=\int_{{\cal M}} dt d^D x\sqrt{g}N\left\{\f{2}{\kappa^2} \left( K^{ij}K_{ij}-\lambda K^{2}\right)-\mathcal{V}[g^{ij},{R^i}_{jkl},\nabla_i]\right\}
\label{action1}
\end{align}
with the higher-spatial-derivative potential $\mathcal{V}$,
satisfying
$[\mathcal{V}]{\leq} D+z$ for the (power-counting)
{renormalizability} \footnote{{There
seems to exist a wide-spread belief
that the original
Ho\v{r}ava action (\ref{action1}) is {\it not} renormalizable due to absence of $a_i=N^{-1} \nabla_i N$-dependent terms, like $a_ia^i$ term, which can be
produced by {\it scalar}-matter-induced loop corrections \cite{DOdo:2015} (see also \cite{Lope:2011,Gros:2021} for different results, {\it i.e,} ``vanishing" coefficient \cite{Lope:2011} and ``different" (non-zero) coefficient value \cite{Gros:2021}). However, the
relevant coefficients should be matter dependent generally and so the effective action
results may or may not affect the genuine gravity sector, depending on the
physics in the matter sector. For example, one might consider a
{\it cancelation} of all matter contributions from fundamental scalars, fermions, and gauge bosons for the $a_ia^i$ term which
can be quite possible by introducing, for example, ``supersymmetry"
(we thank Frank Saueressig for a discussion on this point). So, it is still an open problem
whether the $a_ia^i$ term can be non-zero when all matter contributions are
considered. Moreover, in the ``linear-perturbation" analysis of the purely gravity sector, there has been also claimed that the original Ho\v{r}ava action (\ref{action1}) is inconsistent due to
the singularity (strong-coupling problem) of the $\lambda \rightarrow 1$
limit \cite{Blas:2009}. However, the singularity problem does not occur
in the ``fully non-linear" (constraint) analysis (for example, see the
discussion No. 3 in \cite{Deni:2020}), analogous to the Vainshtein mechanism
in massive gravity \cite{Vain:1972}. Furthermore, the inconsistency
(in the flat Minkowski background) disappears in a more realistic
``time-dependent" background with a {\it inflaton} scalar field, even at
the linear perturbation level (for example, see the discussion No. 2 in
\cite{Shin:2017}). On the other hand, phenomenologically, it is
questionable whether the gravity theory with the $a_ia^i$ term can be a
{\it sensible} theory both astro-physically and cosmologically due to its
important deformations of the Newtonian gravity or GR at
large distances (see \cite{Reye:2010} for a confirmation of GR on large scales) or effects on our currently
accelerating universe which is based on the standard cosmology with a cosmological
constant \cite{Nils:2021}.}},
while keeping only the second-time-derivatives with the terms of $K^{ij}K_{ij}, K^{2}$
in the kinetic part,
through the {\it anisotropic} scaling dimensions, 
${[{\bf x}]=-1, [t]=-z}$ for the dynamical critical exponent $z>1$. Here,
\begin{align}
K_{ij}\equiv \f{1}{2N} \left(\dot{g}_{ij}-\nabla_{i}N_{j}-\nabla_{j}N_{i}\right)
\label{Kij}
\end{align}
is the extrinsic curvature [the overdot $( \dot{~} )$ denotes the
time derivative $\pa_t \equiv \pa_0 =(~)_{,0}$]
and ${R^i}_{jkl}, \nabla_i$ are the Riemann tensor, the spatial covariant derivative for $D$-dimensional
spatial metric $g_{ij}$ on the hypersurface
with the ADM decomposition
\begin{\eq}
ds^2=-N^2 dt^2+g_{ij}\left(dx^i+N^i dt\right)\left(dx^j+N^j dt\right).
\end{\eq}

The peculiar property of the
{HLD} action (\ref{action1}) is that the Lorentz, {\it i.e.}, 
{\it diffeomorphism} ({\it Diff}) symmetry  
in general relativity (GR) is broken into the ``foliation-preserving" diffeomorphism
(${\it Diff}_{\cal F}$)
from either the DeWitt's $\la$ parameter in IR ($\la \neq 1$) \ci{DeWi}, or the higher-derivative terms in UV ($z \geq D$) for power-counting
renormalizability \ci{Lifs,Hora:2009,Viss:2009}. However, it has been unclear how to
{\it canonically} describe the sudden change (reduction) of symmetry beyond the GR limit,
$\la \ra 1, -\mathcal{V} \ra \La +(2/\ka^2) R$, by tracing the missing (Lorentz) symmetry all the way
down to UV.

On the other hand, we have recently found that the constraint structure in the
Hamiltonian formalism looks quite different due to Lorentz-violating terms but ``{\it there exits
the case where we can recover the same number of degrees of freedom as in GR \ci{Deni:2020}, at the fully non-linear level}". This appears to mismatch with the
{\it apparently} broken symmetry in the action and so it suggests some {\it unbroken/enhanced} symmetry {in the Lagrangian formalism} in order to be consistent
with the Hamiltonian formalism.

In this paper, in order to clarify this problem, we examine
the full action symmetry and its associated conservation laws
in Ho\v{r}ava gravity.
Surprisingly, we find that the {\it full}
{\it Diff} symmetry
can be recovered in the Ho\v{r}ava gravity action when the ``super-condition" of
${\cal I}_0 \equiv \nabla_i {\Om}^i=0$ is satisfied,
which exactly corresponds to the {\it tertiary} constraint
in Hamiltonian formalism which is the second-class constraint.
This provides a non-trivial realization of the
{\it Diff} symmetry otherwise
being absent apparently.
From the recovered Lorentz symmetry in the action on the
{\it super-selected} sector, which is still (considered as) an off-shell
condition,
we obtain the
conservation laws with the Noether currents as in covariant theories.
The general formula for the conserved Noether charge reproduces the
mass of four-dimensional static black hole with an {\it arbitrary}
cosmological constant in Ho\v{r}ava gravity, and is independent
of ambiguities associated with the choice of asymptotic boundaries.

{ The organization of the paper is as follows. In Sec. II, we consider the {\it Diff} symmetries of the Ho\v{r}ava gravity in comparison with GR and introduce the {\it super-condition} to consider a {\it super-selected} sector which recovers the Lorentz symmetry in ``off-shell". For the super-selected sector, we obtain the {\it covariant} form of Noether currents. In Sec. III, we consider four-dimensional static black solutions and confirm our general mass formula. In Sec. IV, we discuss several challenging problems, including its implications to Hamiltonian formalism, black hole
thermodynamics, radiations from colliding black holes. In Appendix \ref{app1},
we describe the computational details and Appendix \ref{app2}, we summarize the complete set of constraints in Hamiltonian formalism.}

\section{{\it Diff} Symmetries and Conservation Laws
}

To this end, we start by considering the potential
$\mathcal{V}[g^{ij},{R^i}_{jkl}]$, which is an arbitrary function of
metric $g_{ij}$ and curvature invariants
only, {\it eg.},
\begin{\eq}
-\mathcal{V}= \La+ \xi R+\al R^n +\be \left(R_{ij} R^{ij} \right)^s+\ga ({R^i}_{jkl} {R_i}^{jkl})^r+\cdots,
\label{potential_R}
\end{\eq}
but without (covariant) derivatives, for simplicity.
In order that the construction of a renormalizable action (\ref{action1})
is {\it not} spoiled by the mixing of space and time (derivatives) in the
general coordinate transformations, which could produce ghosts of
higher-time derivatives from the higher-spatial-derivative terms in the potential, we need
to further constrain the allowed coordinate transformations into the foliation-preserving
diffeomorphism $({\it Diff}_{\cal F})$ \ci{Hora:2009},
\begin{align}
\delta_{\xi} t=&-\tilde{\xi}^{0}(t),\quad \delta_{\xi} x^{i}=-\xi^{i}(t,{\bf x}),\label{deltx_Horava}\\
\delta_{\xi} N=&(N\tilde{\xi}^{0})_{,0}+\xi^{k}\nabla_{k}N\label{delN3},\\
\delta_{\xi}{N_{i}}=&\tilde{\xi}^{0}{}_{,0}N_{i}+\xi^{j}{}_{,0}g_{ij}+\nabla_{i}\xi^{j}N_{j}
+N_{i,0}\tilde{\xi}^{0}+\nabla_{j}N_{i}\,\xi^{j}\label{delNi3},\\
\delta_{\xi}{g_{ij}}=&\nabla_{i}\xi^{k}g_{kj}+\nabla_{j}\xi^{k}g_{ki}
+g_{ij,0}\tilde{\xi}^{0}.\label{delg3}
\end{align}

Then, from some straightforward computations \cite{Hora:2009}, one can show
that the
standard action (\ref{action1}) is invariant under ${\it Diff}_{\cal F}$ 
(\ref{delN3}) - (\ref{delg3})
\begin{align}
\delta_{\tilde{\xi}} S
=\int dt d^{D}x \, \left\{ \partial_{t}\big[\tilde{\xi}^{0}(t)\mathcal{L}\big]
+\partial_{i}\big[\xi^{i}({\bf x},t)\mathcal{L} \big]\right\}, \label{action_trans_FDiff}
\end{align}
which reflects the scalar density
nature of the Lagrangian density $\mathcal{L}$, defined by
$S \equiv \int dt d^{D}x \mathcal{L}$. Here, each term in the potential
as well as in the kinetic part of (\ref{action1}) is ``separately" invariant for an arbitrary $\la$
\footnote{For the case $\la = 1/D$,
a separate consideration is needed \ci{Hora:2009}. We will briefly discuss about this in
the discussion No. 4.
}
and all the other parameters in the potential ${\cal V}$
so that
the higher-time-derivative terms, as well as the mixing terms between the time and spatial-derivative terms, from the (Lorentz) transformation of the potential with higher-spatial derivative terms would not occur.

If we consider Einstein-Hilbert (EH) action in GR with $\la=1$ and $-{\cal V} =\La+(2/\ka^2) R$, then there is
an ``{\it accidental}" symmetry enhancement which mixes each term in the action \ci{Park:0910a} so that
we can recover the full {\it Diff} \ci{DeWi}
[$\xi^{\mu} \equiv (\xi^{0},\xi^{i})$],
\begin{align}
\delta_{{\xi}} S_{EH}
=\int dt d^{D}x \, \partial_{\mu}\big[{\xi}^{\mu}
({\bf x},t)\mathcal{L}_{EH}\big] \label{EH_action_trans_Diff}
\end{align}
with
\begin{align}
\delta_{\xi} t=&-\xi^{0}(t,{\bf x}),\quad \delta_{\xi} x^{i}=-\xi^{i}(t,{\bf x}), \label{deltx}\\
\delta_{\xi} N=&(N\xi^{0})_{,0}-N\nabla_{i}\xi^{0}g^{ij}N_{j}+\xi^{k}\nabla_{k}N\label{delN2},\\
\delta_{\xi}{N_{i}}=&\xi^{0}{}_{,0}N_{i}+\xi^{j}{}_{,0}g_{ij}
+\nabla_{i}\xi^{0}\left(g^{kl}N_{k}N_{l}-N^{2}\right)+\nabla_{i}\xi^{j}N_{j}+N_{i,0}\xi^{0}
+\nabla_{j}N_{i}\,\xi^{j}\label{delNi2},\\
\delta_{\xi}{g_{ij}}=&\nabla_{i}\xi^{0}N_{j}+\nabla_{j}\xi^{0}N_{i}+\nabla_{i}\xi^{k}g_{kj}
+\nabla_{j}\xi^{k}g_{ki}+g_{ij,0}\xi^{0}.\label{delg2}
\end{align}

This shows a {\it sudden reduction} of the {\it Diff} symmetry beyond the GR limit in Ho\v{r}ava gravity but it is not quite satisfactory due to lack
of canonical understanding of missing (Lorentz) symmetry. On the other hand,
in the Hamiltonian formalism, 
the symmetries of an action are revealed in the existence of
constraints between the field variables and their conjugate momenta, which being the
{\it canonical} generators of the symmetry transformations. In EH case, there
are $2(D+1)$ {\it first-class} constraints which generate the full {\it Diff}
transformations (\ref{delN2}) - (\ref{delg2}) so that we have {$(D+1)(D-2)/2$}
physical graviton (transverse traceless) modes. Whereas, in HLD case with the action (\ref{action1}),
the constraint structure is quite different, having the {\it second-class}
constraints also due to Lorentz-violating terms, but we have recently found \ci{Deni:2020} that \\

``{\it there exits the case (called Case A)
where
the same number of degrees of freedom can be recovered as in GR, at
the ``fully non-linear" level.}"
\\

This may suggest that, even though its ``apparent" symmetry is just the Lorentz-violating ${\it Diff}_{\cal F}$, {\it the ``full" symmetry of HLD action {in the Lagrangian formalism} can be as large as {\it Diff} in GR}, in order to be consistent with the Hamiltonian analysis!

In order to examine this, which may fill the gap in those two sharply different symmetries
in (\ref{deltx_Horava}) - (\ref{delg3})
and
(\ref{deltx}) - (\ref{delg2}), we study the full {\it Diff}
with an arbitrary $\xi^{\mu} ({\bf x},t)$ to see if the Ho\v{r}ava gravity action
(\ref{action1}) can be invariant in a non-trivial way, just
beyond the {\it apparent} symmetry of {${\it Diff}_{\cal F}$.
To this end in a canonical way, we start by considering
 the variation of the action (\ref{action1}) with the potential $\mathcal{V}[g^{ij},{R^i}_{jkl}]$, under
the arbitrary variations of the ADM variables,
\begin{align}
\delta {S}=\int dt d^D x \left[-{\cal H} \delta N -{\cal H}^i \delta N_i+{\bf E}^{ij}\delta g_{ij}+\partial_{\mu} \Theta^{\mu} (\delta N_i,\delta g_{ij}) \right],
\label{del_action_general}
\end{align}
where we have the bulk terms with
\begin{align}
{\cal H}\equiv&-\frac{\delta S}{\delta N}=\sqrt{g}\left[\left(\frac{2}{\kappa^2}\right)
\left(K_{ij}K^{ij}-\lambda K^{2}\right)+{\cal V}
\right],\label{E}\\
{\cal H}^i \equiv&-\frac{\delta  S}{\delta N_i}=\, - 2\sqrt{g}\, \left(\frac{2}{\kappa^2}\right) \nabla_{j}\left(K^{ij}-\lambda g^{ij} K\right),\label{E1}\\
{\bf E}^{ij}\equiv& \frac{\delta  S}{\delta g_{ij}}
= E^{ij}_{(0)}-\sqrt{g}\left[ NP^{iklm}R^{j}{}_{klm}+\frac{1}{2}N g^{ij}\mathcal{V}[g^{ij},R^{i}{}_{jkl}]
-2\nabla_{k}\nabla_{l}(NP^{iklj})\right],
\label{E_ij_general}
\end{align}
and the boundary terms $\Theta^{\mu}  (\delta N_i,\delta g_{ij})$ with
\begin{align}
\Theta^{0} \equiv &\,
\sqrt{g} \left(\frac{2}{\kappa^2}\right)\left(K^{ij}-\lambda g^{ij} K\right) \delta g_{ij},
\label{Theta^0} \\
\Theta^{i}   \equiv & \sqrt{g} \left(\f{2}{\ka^2} \right) \left(2N^{l}{G}^{ijkm}K_{km}\delta g_{jl}-N^{i}{G}^{ljmn}K_{mn}\delta g_{jl}-2{G}^{kjil}K_{kj}\delta N_{l}\right) \no \\
&+2 \sqrt{g} P^{jkil}N\nabla_{k}\delta g_{lj}-2 \sqrt{g} \delta g_{lj}\nabla_{k}(P^{jikl}N),
\label{Theta_i_general}
\end{align}
where $G^{ijkm} \equiv \de^{ijkm}-\la g^{ij} g^{km}$ is the DeWitt metric \cite{DeWi}.
Here, the tensor
\begin{align}
P_{i}{}^{jkl}\equiv \left(\frac{\partial {\cal L}}{\partial R^{i}{}_{jkl}}\right)_{g^{mn}}=-\left(\frac{\partial {\cal V}}{\partial R^{i}{}_{jkl}}\right)_{g^{mn}},
\label{P}
\end{align}
by treating $g^{ij}$ and $R^{i}{}_{jkl}$ are independent fields \cite{Wald:1994},
has the same symmetries in the indices as those of Riemann tensor ${R^i}_{jkl}$
({see Appendix \ref{app1} for} the explicit forms for ${\bf E}^{ij}$ and $\Theta^i$).

Plugging
{\it Diff} transformations (\ref{delN2}) - (\ref{delg2})
into the arbitrary variation
(\ref{del_action_general}) and doing some straightforward computations, we obtain
the action transformation \cite{Kiri:2011},
\begin{align}
\delta_{\xi}{S}&=\int dt d^{D}x \left[-{\cal H}\delta_{\xi} N-{\cal H}^{i}\delta_{\xi} N_i+{\bf E}^{ij}\delta_{\xi} g_{ij}+\partial_{\mu} \Theta^{\mu} (\delta_{\xi} N_i,\delta_{\xi} g_{ij})\right] \label{variation_0}\\
&=\int dt d^{D}x \left[\xi^{0}{\cal I}_{0}+\xi^{i}{\cal I}_{i}+\partial_{\mu} \Psi^{\mu} (\delta_{\xi} N_i,\delta_{\xi} g_{ij})\right],\label{variation_1}
\end{align}
where
\begin{align}
{\cal I}_{0}\equiv& N\dot{\cal H }
-\nabla_{m}(NN^{m}{\cal H })+N_{i}\dot{\cal H}^{i}
+\nabla_{m}\left[{\cal H}^{m}(g^{jl}N_{j}N_{l}
-N^{2})\right]+\dot{g}_{ij}{\bf E}^{ij}-2\nabla_{m}(N_{i}{\bf E}^{mi}) ,
\label{I0_general}\\
{\cal I}_i\equiv&
(g_{ij} {\cal H}^j
)_{,0}
+\nabla_{m}({\cal H}^{m}N_{i})
-{\cal H }\nabla_{i}N-{\cal H}^{j}\nabla_{i}N_{j}-2g_{ij}\nabla_{m}{\bf E}^{jm},
\label{Ik_general}\\
\Psi^0\equiv&-\xi^{0}\left( N {\cal H }+{N}_{i}{\cal H}^{i}\right)-\xi^{j}g_{ij}{\cal H}^{i} +\Theta^0, \label{Omega0_general}\\
\Psi^i\equiv&\xi^0 \left[NN^{i}{\cal H }-{\cal H}^{i}(g^{lj}N_{l}N_{j}
-N^{2})+2 N_{j}{\bf E}^{ij} \right] +\xi^j \left( -N_j {\cal H}^{i}+2 g_{jl} {\bf E}^{il}\right)+\Theta^i_.  \label{Omega_i_general}
\end{align}
Here,
it is important to note that the formal expressions of
(\ref{I0_general}) - (\ref{Omega_i_general}) are generally valid for arbitrary
potential $\mathcal{V}[g^{ij},{R^i}_{jkl}]$ though the explicit expressions of
${\cal H}, {\cal H}^i, {\bf E}^{ij}, \Theta^\mu, \Psi^{\mu}$ may depend on the
potential form.

Now, by expressing the time-derivatives in terms of extrinsic curvature $K_{ij}$ in (\ref{Kij}) and
using the definitions (\ref{E}) -
(\ref{E_ij_general}), {\it but without using the dynamical equations of motion ${\bf E}^{ij}=0$, nor the constraints ${\cal H}\approx 0, {\cal H}^i \approx 0$}, we obtain
\begin{\eq}
{\cal I}_{0} &=&\nabla_{i}\left\{2N^{2}\left[\nabla_{j}\pi^{ij}+\left(\f{\ka^2}{2} \right) \left( \frac{2\lambda}{\lambda D-1}\left(\pi\nabla_{l}P^{kl}{}_{k}{}^{i}-P^{kl}{}_{k}{}^{i}\nabla_{l}\pi\right)
+2P_{jkl}{}^{i}\nabla^{k}\pi^{jl}
-2\pi^{jl}\nabla^{k}P_{jkl}{}^{i} \right) \right]\right\} \no \\
&\equiv &\nabla_{i}{\Omega}^{i}, \label{I0_final} \\
{\cal I}_i & = & 0, \label{Ik_final}\\
\Psi^{0}&=&\xi^{0}\mathcal{L}+
\partial_{i} { \cal U}^{0i}, \label{Omega0_final}\\
\Psi^{i}&=&\xi^{i}\mathcal{L}+{\bf \Sigma}^{i}+\partial_{0} {\cal U}^{i0}+\partial_{j} {\cal U}^{ij}, \label{Omega_i_final}
\end{\eq}
where ${\bf \Sigma}^{i}$ and ${\cal U}^{\mu \nu}=-{\cal U}^{\mu \nu}$, which is called the ``super-potential" \cite{Freu:1939,Berg:1949}, are given by
\begin{\eq}
{\bf \Sigma}^{i} &=&
2N^{2}\Bigg[\left(\f{\ka^2}{2} \right)\left( \frac{2\lambda}{\lambda D-1}
\left(P^{li}{}_{lk}\nabla^{k}(\xi^{0}\pi)-\xi^{0}\pi\nabla^{k}P^{li}{}_{lk}\right)
+2\xi^{0}\pi^{jl}\nabla^{k}P_{jkl}{}^{i}-2P_{jkl}{}^{i}\nabla^{k}(\xi^{0}\pi^{jl}) \right)\no \\
&+&\pi^{ij}\nabla_{j}\xi^{0}-\xi^{0}\nabla_{j}\pi^{ij}
\Bigg], \label{Sigma_i}\\
{\cal U}^{0i}&=&-{\cal U}^{i0} = 2 \sqrt{g} (\xi^0N_j +\xi_j)  \left(\frac{2}{\kappa^2}\right) \left( K^{ij}-\la g^{ij}K \right),\\
{\cal U}^{ij}&=&-{\cal U}^{ji} \label{U_ij}
\end{\eq}
({see} Appendix \ref{app1} for the computational details of
(\ref{I0_final})
- (\ref{Omega_i_final}) and the explicit forms for 
${\cal U}^{ij}$ of (\ref{U_ij})). Here, we note that ${\cal I}_i=0$
identically, which is the analogue of the spatial component of the (contracted)
Bianchi identity $\hat{\nabla}_{\mu} G^{\mu i} = 0$ in GR, for the
Einstein tensor $G^{\mu \nu} = \hat{R}^{\mu \nu}-(1/2)\hat{g}^{\mu \nu} \hat{R}$
with the $(D+1)$-dimensional Ricci tensor $\hat{R}^{\mu \nu}$, Ricci
scalar $\bar{R}$ \cite{Ande:1998} and covariant
derivative $\hat{\nabla}_{\mu}$: {In the Hamiltonian formalism, it}
is due
to the first-class nature of momentum constraint ${\cal H}_i \approx 0$, the
generator of the spatial ${\it Diff}$, even in HLD gravity \cite{Deni:2020}.
In ${\cal I}_{0}$, we have replaced the extrinsic curvature
with the canonical momenta
$\pi_{ij}=(2/\kappa^2)\sqrt{g} (K_{ij}-\lambda K g_{ij})$
in order to compare { it} with
the Hamiltonian analysis.


In GR case,
we have $
{\bf \Sigma}^i=0$ and
${\cal I}_{0}
= 0$, which corresponds to the time-component of
Bianchi identity, $\hat{\nabla}_\mu G^{\mu 0}=0$ and hence obtain the usual Lorentz invariance of
(\ref{EH_action_trans_Diff}) from (\ref{variation_1}).
For the ${\it Diff}_{\cal F}$ in Ho\v{r}ava gravity case,
the parameter $\xi^0(t)\equiv \tilde{\xi}^0 (t)$ for the $\xi^{0}{\cal I}_{0}$ in (\ref{variation_1})
can be factorized out from the space integration and the $\xi^{0}{\cal I}_{0}$
term turns into a (spatial) boundary term, but it is ``exactly canceled" by another
boundary term $\nabla_i {\bf \Sigma}^{i}$
so that we
can obtain the invariance
of (\ref{action_trans_FDiff}) under ${\it Diff}_{\cal F}$.

For the full {\it Diff} with an arbitrary $\xi^0 ({\bf x},t)$, on the other hand,
the {${\cal I}_{0}
$ term, which is {\it non-zero} (see the explicit expression in (\ref{I0_explicit})) unless we consider $\lambda = 1$ and vanishing of all the higher-derivatives terms (as in the GR case),}
can not be removed from the bulk terms anymore and may
result the non-invariance of the HLD action generally,
confirming the usual belief of its generic Lorentz violation. Therefore, the only way
{ of}
obtaining the full {\it Diff} invariance
{ for}
the HLD action, {\it if it exits},
would be to consider
a ``super-condition",
\begin{\eq}
{\cal I}_{0}\equiv \nabla_{i} {\Om}^i=0,
 \label{I0_constraint}
\end{\eq}
which {\it super-selects} the Lorentz-invariant sector in HLD gravity.
{We note also, from (\ref{E1}) and (\ref{I0_final}),}
that the super-condition can be written as
\beq
{\cal I}_{0}= {\Om}- \nabla_i (N^2 {\cal H}^i)=0,
\eeq
where
\beq
{ \Om} \equiv
\nabla_i (N^2 C^i)
\label{I0_constraint_2}
\eeq
{ with
\beq
C^i \equiv {\ka^2}  \left( \frac{2\lambda}{\lambda D-1}\left(\pi\nabla_{l}P^{kl}{}_{k}{}^{i}-P^{kl}{}_{k}{}^{i}\nabla_{l}\pi\right)
+2P_{jkl}{}^{i}\nabla^{k}\pi^{jl}
-2\pi^{jl}\nabla^{k}P_{jkl}{}^{i} \right).
\label{C}
\eeq
Then, it is remarkable that (\ref{I0_constraint_2})}
reduces to
the {\it tertiary} constraint ${\Om}
\approx 0$ in Hamiltonian formalism {\cite{Deni:2020},}
using the dynamical equations of motion
\footnote{$\dot{ g}_{ij}= \{g_{ij}, H_C \},~\dot{\pi}^{ij}=\{\pi^{ij}, H_C \}$
with the canonical Hamiltonian
$H_C=\int dx^D \left\{ N {\cal H} +N_i {\cal H}^i\right\}$.
},
{\it i.e.}, in {\it on-shell}, from
${\dot{\cal H}}\approx {\Om}/N \approx 0$,
with the Hamiltonian constraint ${\cal H}\approx 0$ and
the momentum constraint ${\cal H}^i \equiv -2 \nabla_j \pi^{ij} \approx 0$
{
(see Appendix {B} for a summary of the complete set of constraints)}.
In other words, the super-condition (\ref{I0_constraint}) { in our
Lagrangian formalism of HLD gravity} is {\it on-shell} equivalent to the tertiary constraint in
{the Hamiltonian formalism. On the other hand, in GR, the tertiary
constraint}
is trivial},
up to the Hamiltonian and momentum constraints, ${\cal H}\approx 0, {\cal H}^i \approx 0$, which are the secondary constraints. \footnote{From this result,
one can use (\ref{I0_general}) as
an off-shell, Lagrangian definition of the {\it tertiary} constraints via the
terms of $N\dot{\cal H }
$ and $N_{i}\dot{\cal H}^{i},
$ which are usually quite cumbersome in Hamiltonian formalism. Actually, using
the new definition in this paper we have generalized the Hamiltonian analysis
on tertiary constraint for $\dot{\cal H }
$ in \cite{Deni:2020}, where we have considered only an arbitrary function of
$R$ in the potential (\ref{potential_R}) {
(see also Appendix {B})}.
Moreover, one can find easily
the exactly same constraint algebra for the Hamiltonian and momentum constraints as in the
previous case \cite{Deni:2020},
 $
 \{ \left<\eta {\cal H} \right>,\left<\zeta {\cal H}\right> \}=
 \left<  \left( \eta \nabla_i \zeta-\zeta \nabla_i \eta \right) C^i \right> ,
 \{ \left< \eta {\cal H} \right>,\left< \zeta^i {\cal H}_i\right>\}
 =- \left<  \zeta^i \nabla_i \eta {\cal H} \right>  ,
 \{ \left< \eta^i {\cal H}_i \right>,\left< \zeta^j {\cal H}_j \right> \}
 = \left< \left( \eta^i \nabla_i \zeta^j-\zeta^i \nabla_i \eta^j \right) {\cal H}_j \right>  ,
$
for { the {\it Diff} parameters $\eta, \zeta$, and} the smeared constraints, $\left< \eta  {\cal H} \right> \equiv \int d^D x~ \eta {\cal H}$, {\it etc.}
}
{The}
{intimate relation to constraints {in the
Hamiltonian formalism}} may
{support for} our introduction of the super-condition (\ref{I0_constraint}) {in the Lagrangian formalism}.
{However,}
it is important to note that{, in our Lagrangian formalism,} the super-condition is {\it assumed} to be valid
even {\it off-shell}
\footnote{From the form of the Bianchi identity in GR,
$\bar{\nabla}_\mu {G^{\mu}}_{ 0}=\bar{\nabla}_0 {G^{0}}_{ 0}
+\bar{\nabla}_i {G^{i}}_{0}=\left( {\cal H}_{GR}/2 \sqrt{g} \right)_{,0}
+ \cdots+\nabla_i \left( {\cal H}^i /2N \sqrt{g} \right )=0$ with
${G^{0}}_{ 0}= {\cal H}_{GR}/2 \sqrt{g},~ {G^{i}}_{0}={\cal H}^i /2\sqrt{g}N $
for the Hamiltonian
constraint in GR, ${\cal H}_{GR}\approx 0$,
the super-condition suggests {\it formally} the same Bianchi identity but now with ${G^{0}}_{ 0}\equiv {\cal H}/2 \sqrt{g}$ in HLD
gravity also. This provides a more fundamental {\it off-shell} reason for the appearance of the tertiary constraint via $\dot{\cal H}$.}, {\it i.e.} prior to
considering the ``classical geometry dynamics" which extremises the action.
In this way, the {\it full} {\it Diff} can be maintained all the way down to UV, even beyond
the GR limit.
Here, the super-condition
${\cal I}_0 = 0$ in HLD gravity
corresponds to
the temporal component of Bianchi identity, $\hat{\nabla}_{\mu} G^{\mu 0}= 0$ in GR.



Finally, from (\ref{del_action_general}), (\ref{variation_1}), and with the help of the super condition ${\cal I}_0 = 0$ and the Bianchi identity ${\cal I}_i = 0$, we can obtain the
Noether currents ($\Sigma^0=0$),
\begin{\eq}
{\bf \cal{J}}^{\mu}(\delta_{\xi} g) &\equiv& \Theta^{\mu} (\delta_{\xi} N_i,\delta_{\xi} g_{ij})-\Psi^{\mu} (\delta_{\xi} N_i,\delta_{\xi} g_{ij})\no \\
&=& \Theta^{\mu} -\xi^{\mu} {\cal L}-{\bf \Sigma}^{\mu}-\partial_{\nu} {\cal U}^{\mu \nu},\label{Noether}
\end{\eq}
which satisfies
\begin{align}
\partial_{\mu} {\bf \cal{J}}^{\mu}(\delta_{\xi} g) ={\cal H}\delta_{\xi} N+{\cal H}^{i}\delta_{\xi} N_i-{\bf E}^{ij}\delta_{\xi} g_{ij}. \label{Noether_conserv}
\end{align}
Note that the
Noether currents satisfies the usual conservation laws {\it on-shell}, {\it i.e.} ${\cal H}={\cal H}^{i}={\bf E}^{ij}=0$, for an arbitrary {\it Diff} transformation, but {\it off-shell} for Killing vectors $\xi^{\mu}$, $\delta_{\xi} N=\delta_{\xi} N_i=\delta_{\xi} g_{ij}=0$. The second term $\Sigma^{\mu}$ is due to the {\it apparent} non-covariance of the Horava action.\footnote{The non-covariance term appears also in Chern-Simons theories \cite{Oh}.} Moreover, the last part in the Noether current (\ref{Noether}), corresponds to the identically conserved or {\it off-shell current} ${\cal J}^{\mu}_{\rm off} \equiv \partial_{\nu} {\cal U}^{\mu \nu}$ \cite{Freu:1939,Berg:1949,Kim:2013}. Then, the conserved charge passing through a hypersurface $\Sigma$ is given by
\begin{\eq}
Q(\xi) = \int_{\Sigma} d^{D}x \sqrt{g} ~n_{\mu }{J}^{\mu}(\delta_{\xi} g)
\label{Noether_charge}
\end{\eq}
for the unit normal vector $n^{\mu}$ of $\Sigma$ and the {\it covariantly} conserved charge $J^{\mu}=(\sqrt{g} N)^{-1} {\cal J}^{\mu}$, satisfying $\nabla_{\mu} J^{\mu}=0$ \cite{Carr:2004} and the physically {\it measurable} charge can be obtained by subtracting the background charge $\bar{Q} (\xi) \equiv \int_{\Sigma} d^{D}x \sqrt{\bar{g}} ~\bar{n}_{\mu } \bar{ {J}}^{\mu}(\delta_{\xi} \bar{g})$, {\it i.e.}, $ Q(\xi)_{\rm phys} \equiv Q(\xi)-\bar{Q}(\xi)$ generally, where the bars denote the background quantities.

\section{An Example: Static Black Holes in $(3+1)$ Dimensions}

In order to check the general Noether charge formula (\ref{Noether_charge}), let us consider the static metric ansatz
\begin{\eq}
ds^2=-N^2(r) dt^2+\f{dr^2}{f(r)}+r^2 d \Om_k^2
\end{\eq}
by which the original Horava gravity action in $(3+1)$ dimensions \cite{Hora:2009} reduces to the case with the potential form of (\ref{potential_R}), due to the vanishing Cotton tensor, $C^{ij}\equiv \ep^{ikl} \nabla_k ({R^j}_l-{\de^j}_l R/4)=0$. Here, $d \Om_k^2$ denotes the line element for two-dimensional surface with a constant scalar curvature, $R^{(2)}=2k$
for spherical, plane, and hyperbolic topologies
with $k=+1,0,-1$, respectively.

Then, for the time-like Killing vector $\xi^{\mu}=(1,0,0,0)$ and the normal vector $n_{\mu}=(-N,0,0,0)$, the only non-vanishing contributions in the Noether charge (\ref{Noether_charge}) come from the second term in the
current (\ref{Noether}) and, after the angular integrations, is given by
\begin{\eq}
Q(\xi^0)&=&\Om_k \int^r_0 dr r^2 \left( \f{N}{\sqrt{f}}\right)
\xi^0 {\cal L} \label{Q_Horava}\\
&=&-{\si }~  \xi^0 \left( \f{N}{\sqrt{f}}\right)\left[-\f{\la}{r}(f-k)^2+2 (\om-\La_W) r (f-k)-\La_W^2 r^3 \right] \Big|^{r} \no \\
&+&{\si}  \int^r_0 dr~ \partial_r \left( \xi^0  \f{N}{\sqrt{f}}\right) \left[-\f{\la}{r}(f-k)^2+2 (\om-\La_W) r (f-k)-\La_W^2 r^3 \right] \no \\
&-&{\si}~ (\la-1)  \int^r_0 dr \xi^0 \left( \f{N}{\sqrt{f}}\right) \left[ \f{(f-k)^2}{r^2}+\f{(\pa_r f)^2}{2}\right] \no
\end{\eq}
in the usual { parametrization,
\beq
\xi \equiv \f{\ka^4 \mu^2 ( \La_W +\om)}{8 (1-3 \la)}, ~
\al \equiv \f{\ka^2 \mu^2 (1-4 \la)}{32 (1-3 \la)},~\beta \equiv \f{\ka^2 \mu^2 }{8},~\ga \equiv 0,~
 \La \equiv-\f{2 \ka^2 \mu^2 \La_W^2}{8 (1-3 \la)}
\eeq}
with ${\si} \equiv \Om_k \ka^2 \mu^2/8(3 \la-1)$, the IR-modification parameter $\om$, and $D$-dimensional cosmological constant parameter $\La_W$ \cite{Hora:2009,Lu:2009,Keha:2009,Park:2009,Cai:2009,Argu:2015}. Here, it is important that we need to (1) first, change the Lagrangian into the total derivatives (the second line in the above formula) plus the bulk terms (the third and fourth lines), and then (2) second, compute the charge by plugging the known solutions: If we first plug the solutions into the Noether charge formula (\ref{Q_Horava}), we obtain the trivially vanishing charge because the Lagrangian in the charge formula is proportional to the Hamiltonian constraint ${\cal H} \approx 0 $, which is solved by the solutions.\footnote{This looks tricky but this kind of prescription seems to be essential to get the right answer {(see also} \ci{Carr:2004} for some related discussions). In particular, for $\la=1$ and $(N/\sqrt{f})=constant$, our charge (\ref{Q_Horava}) agrees with the (generalized) Misner-Sharp mass \cite{Cai:2009}.}

Now, by plugging the general static vacuum solution for arbitrary cosmological constant parameter $\La_W$ and IR parameter $\om$ with $\la=1$ \cite{Park:2009,Argu:2015}, whose {\it uniqueness} is guaranteed by the corresponding {\it Birkhoff's theorem} \ci{Deve:2018} (for more general cases, see the discussion No. 8 below),
\begin{\eq}
N^2=f=k+(\om-\La_W)r^2 +\ep \sqrt{r [\om (\om-2\La_W)r^3+{{\bm \beta}}]},
\label{Park_sol}
\end{\eq}
where $\ep=\pm 1$ and ${{\bm \beta}}$ is an integration constant
\footnote{$\ep=-1 ~(+1)$ represent an asymptotically flat or anti-de Sitter
(de Sitter) with $\om, \mu^2>0$ ($\om, \mu^2<0$)
\ci{Argu:2015}. {Here, we consider only the {\it GR-branch} solutions
which have the GR limits in IR regime as in \ci{Lu:2009,Keha:2009,Park:2009}.
The other choices of the $\ep$ represent the {\it non-GR branch} solutions which
do not have the GR limits and these are important for studying the genuine
higher-derivative solutions
 \ci{Kim:2015}.}},
we can obtain
\begin{\eq}
Q(\xi^0) ={\si} {{\bm \beta}},
\label{Q_Horava_sol}
\end{\eq}
which exactly agrees with the mass in the conventional Hamiltonian approach \ci{Cai:2009,Donn:2011,Bell:2011}. Note that the mass ${\cal M} \equiv Q(\xi^0) $ satisfies the first law of black hole thermodynamics
\begin{\eq}
\delta {\cal M}=T_H \delta {\cal S}
\label{1st_law}
\end{\eq}
with the black hole
temperature $T_H$ \footnote{{See \ci{Liu:2012}, for an explicit computation of the Hawking radiation and temperature for {\it relativistic} matters, based on the quantum tunneling approaches.} } and the entropy ${\cal S}$ with a logarithmic term, up to an arbitrary constant ${\cal S}_0$,
\begin{\eq}
T_H &\equiv& \f{\hbar \kappa|_{H}}{2 \pi}
=\f{\hbar (3 \La_W^2 r_H^4+2 k (\om-\La_W) r_H^2-k^2)}{8 \pi r_H (k+ (\om-\La_W)r_H^2)}, \\
{\cal S}&=&\f{4 \pi {\si} }{\hbar} ((\om-\La_W) r_H^2+2k~ {\rm ln} r_H)+{\cal S}_0
\end{\eq}
for the surface gravity $\kappa_H=(1/{2 }) \pa_r f|_{r_H}$ at the black hole horizon $r_H$.

Two remarkable properties of this result are as follows. First, the result
(\ref{Q_Horava}) does not depend on the boundary $(D-1)$-hypersurface only if there
is a time-like Killing vector inside the boundary.
This means that the boundary
needs not to be an asymptotic infinity even in asymptotically de-Sitter
space as well as in flat or anti-de Sitter space (for similar results in
covariant theories, see \ci{Deru:2004,Dola:2018}). Second, related to the
first property, there are no divergences in anti-de Sitter space, and it is
independent on the ambiguities associated with the choice of asymptotic boundary
at $r\ra \infty$ in de Sitter space \ci{Park:1998}. So, for the asymptotically
de Sitter black hole, the boundary can be any region between the outer
black hole horizon $r_+$ and the cosmological horizon $r_{++}$, {\it i.e.}
$r_+ < r < r_{++}$.


\section{Concluding Remarks}{\label{section4}}

In conclusion, we have shown that the Lorentz symmetry, which is represented by {\it Diff} symmetry, is preserved on the super-selected sector of ${\cal I}_0 \equiv \nabla_i {\Om}^i=0$ even in HLD gravity action where the explicit Lorentz violating terms are introduced
for (power-counting) renormalizablity.
This indicates that the full {\it Diff} symmetry of HLD action can be as large as the {\it Diff} in GR and, from the obtained full {\it Diff} symmetry, we find the conservation laws with the Noether currents as in covariant theories \cite{Berg:1949}. Several further remarks about challenging problems are in order.\\

1. The super-condition ${\cal I}_0 \equiv \nabla_i { \Om}^i=0$ is similar to the Maxwell's equation $\nabla_i {B}^i=0$ for the magnetic field ${B}^i$ without (magnetic) monopoles. If we define the $(D-2)$-form "currents'' for $ D$-dimensional space with the component
${J}^{i...n}=\epsilon^{i...njk} \nabla_j {\Omega}_k$, it
satisfies the (spatial) conservation laws $\nabla_i J^{i...n}=0$ as in the equations ${j}^i=\ep^{imn} \pa_m {B}_n$ for the magneto statistics with the electric currents ${j}^i$. Then, we can solve ${\Om}^i$ in terms of the currents ${J}_{i...n}$, which are the additional data for a complete specification of ${ \Om}^i$. For example, if the super-condition $\nabla_i {\Om}^i=0$ and $\epsilon^{i...njk} \nabla_j {\Omega}_k={J}^{i...n}=0$ are satisfied for the {\it whole-space} region, {\it i.e.} without singularities, then ${\Om}^i=0$ would be the only solution and this would correspond to ``Case A" in the Hamiltonian analysis of \cite{Deni:2020} where the degrees of freedom in HLD gravity are the same as in GR at the fully non-linear level. Otherwise, ${\Om}^i$ would be non-vanishing generally due to either (a) non-trivial tolopoly/cohomology, or (b) singularities, or from (c) non-vanishing current ${J}^{i...n} \neq 0$. This latter case would correspond to ``Case B" in \cite{Deni:2020} where an extra scalar graviton mode exists in Hamiltonian analysis of HLD gravity. It would be interesting to find the generic (formal) solution of ${\Om}^i$ in curved space, corresponding to Biot-Savert's law in electromagnetism in Minkowski space-time.
\\

2. From the invariance of the action (\ref{variation_1}), we have obtained the
tertiary constraints $\nabla_i ({N^2 C}^i)=0$ in Hamiltonian formalism via
$\dot{\cal H }
$ and $\dot{\cal H}^{i}
$ in the super-condition ${\cal I}_0=0$. As have been noted above, the action invariance does not necessarily mean the same degrees of freedom as in GR, which is manifestly Lorentz invariant. In Hamiltonian formalism, we need to find a complete set of constraints to completely specify the degrees of freedom. This implies that we need more consistency analysis in Lagrangian formalism, corresponding to the preservation of constraints in Hamiltonian formalism. We suspect that the higher-order invariance of $\delta_{\xi} \delta_{\eta} ...\delta_{\zeta} S=0$ would be important in HLD gravity and needs to be considered in order to obtain the complete set of constraints, consistently with the Hamiltonian formalism.
\\

3. Our formulation about the gravity sector is self-contained and independent on the matter sector. If we now consider matter action $S_m$
as well, which may have non-relativistic higher-derivative terms also in accordance with the HLD gravity, the additional contributions to the action transformation are
$\int dt d^D x [\xi^0 \sqrt{\hat{g}} \hat{\nabla}^{\mu} T_{\mu 0}+\xi^i \sqrt{\hat{g}} \hat{\nabla}^{\mu} T_{\mu i}]$ with the energy-momentum tensors for matters $T_{\mu \nu} =-\f{2}{\sqrt{\hat{g}}} \f{ \de S_m}{\de \hat{g}^{\mu \nu}}$ , for the $(D+1)$-dimensional metric $\hat{g}_{\mu \nu}$ and its associated covariant derivatives $\hat{\nabla}^{\mu}$,
together with the matter contributions to the boundary terms $\Theta^{\mu}$ and $\Psi^{\mu}$.
But, from the
super-condition ${\cal I}_0=0$ and the Bianchi identity ${\cal I}_i=0$ in
the gravity sector, which indicating their geometrical origin,
the consistent theory with the full {\it Diff} is possible only for the covariantly-conserved matters, {\it i.e.} $\hat{\nabla}^{\mu} T_{\mu \nu}=0$, regardless higher-derivatives in HLD gravity. In other words, only the energy-conserving matters $\hat{\nabla}^{\mu} T_{\mu 0}=0$, as well as the momentum-conserving matters $\hat{\nabla}^{\mu} T_{\mu i}=0$, can be consistent with the HLD gravity. Actually, there seems to exist some evidence for the covariant conservation laws of matter's energy momentum tensors and of the effective energy-momentum tensors from the higher-derivative terms, separately, for spherically symmetric case \ci{Son:2010}.\footnote{One can {\it formally} write the Horava gravity's equations of motion into a {\it covariant} Einstein's equation $ G^{\mu \nu}=8 \pi G  T_{\rm{eff}}^{ \mu \nu}$ by considering the higher-derivative contributions as the effective energy-momentum tensor $T_{\rm{eff}}^{ \mu \nu}$. Then, from the usual (covariant) Bianchi identity on the Einstein tensor $\hat{\nabla}_{\mu} G^{\mu \nu}=0$, one can find the covariant conservation laws $\hat{\nabla}_{\mu}T_{\rm{eff}}^{ \mu \nu}=0$ \ci{Son:2010}. However, the geometric origin of this identity/property is still unknown.} It would be interesting to see whether the covariant form of the conservation laws holds generally, as another super-condition in matter sectors.\\

4. For the special value of IR Lorentz-violation parameter $\la = 1/D$, the theory has {\it anisotropic} Weyl symmetry
additionally \ci{Hora:2009} but a separate consideration is needed.
 Based on the Hamiltonian analysis \cite{Deni:2020}, which gives the same degrees of freedom as in GR, its full symmetry would be also as large as that of GR, though its details of the symmetry are different. It would be interesting to clarify the full action symmetry also and its connection to the Case A.\\

5. From the obtained Lorentz symmetry of HLD action for the super-selected sector of
$\nabla_i \Om^i=0$,
one can consider the corresponding
{\it Ward-like}
identity $\left< \de_{\xi}{\cal F} \right> -(i/\hbar) \int \xi \left< {\cal F} \partial_{\mu} {\cal J}^{\mu} \right>  dt d^D x = 0 $
for a {\it Diff} invariant observable ${\cal F}$,
from the {\it Diff} invariance of the path-integral measure with the first and second-class constraints \cite{Senj:1976}. The proof of renormalizability for HLD gravity from the gravitational Ward-like identity would be a challenging problem. \\

6. In our formulation, we have considered the arbitrary potential $\mathcal{V}[g^{ij},{R^i}_{jkl}]$ without derivatives $\nabla_i$.
The inclusion of derivatives in the potential, {\it i.e.}
$\mathcal{V}[g^{ij},{R^i}_{jkl},\nabla_i]$ would be more desirable to describe
the most general systems as in the action (\ref{action1}). The computations are straightforward but, due to
the complications, we have not yet succeeded in obtaining the canonical form of
(\ref{variation_1}). However, we believe that the formulation itself should not
depend on the existence of derivatives in the potential so that there should
be no fundamental problem to get the corresponding canonical forms.\\

7. The {\it Diff} invariance of HLD gravity sheds a new light on the very
meaning of black hole entropy and its thermodynamical laws, due to the
revival of Lorentz invariant concept of the event horizon, which has been
essential to give an absolute meaning to black hole entropy as a
{\it measure of observable ignorance} inside the event horizon as well
as the role of the {\it universal horizons} in the presence of the
(Lorentz-invariant) event horizons \cite{Blas:2011}
\footnote{ In Lorentz-violating gravities, the thermodynamical
properties, like the Hawking radiation have been long-standing issues and
there have been some controversial results. In \cite{Berg:2012,Crop:2013},
it is argued of the radiations at the universal horizon but {\it none}
at the Killing horizon. In \cite{Mich:2015},
it is shown the opposite results in a more direct calculation, {\it i.e.}, the radiations at the Killing horizon but none at the universal horizon, which seems to support for our new formulation and some earlier results on Hawking radiations for relativistic matters, like \ci{Peng:2009,Liu:2011,Liu:2012}. Recently, it has been clarified that the disagreements were due to the different choices observer's frames (or vacuum) \ci{Herr:2020}.}. In particular, for the $k=1$ black holes in an asymptotically flat/AdS space, the logarithmic correction to the usual Bekenstein-Hawking
entropy implies the ``positive" {\it minimum} of horizon radius $r_{H}$ for the
{\it positive} black hole entropy ${\cal S}$, which is consistent
with the existence of a positive minimum for the mass ${\cal M}$ where the Hawking
temperature $T_H$ vanishes \ci{Argu:2015}.\footnote{One can choose $S_0$ so that the two minimum horizon radii agree.
This choice achieves the third law of black hole thermodynamics, {\it i.e.},
${\cal S}=0$ at $T_H=0$.} Moreover, in that case, the black hole entropy ${\cal S}$ increases (the second law of black hole thermodynamics) by $\Delta {\cal S}=({8 \pi {\si}}/{\hbar}) [(\om-\La_W) r_H+2k/ r_H] \Delta r_H=T_H \Delta {\cal M}$ for the increased mass $\Delta {\cal M}>0$. From the associated increase of area \footnote{This indicates the energy conditions, especially the weak energy conditions (WEC) and the null energy conditions (NEC) are not violated by the higher-derivative contributions to the (effective) energy-momentum tensors. For example, for the asymptotically flat case, {\it i.e.}, $\La_W=0$, see \ci{Son:2010}.} $A_H=4 \pi r_H^2$, one can compute the upper bound to the energy of the gravitational radiations when one black hole captures another. For the asymptotically flat black holes ($\La_W=0$) \ci{Keha:2009} with the area (${{\bm\be}} \equiv 4 \om M$)
\begin{\eq}
A_H=8 \pi M^2 [1+(1-(2 \om M^2)^{-1})^{1/2}-(4 \om M^2)^{-1}],
\label{area}
\end{\eq}
the increased area gives the inequality ($m \equiv \om^{1/2} M$)
\begin{\eq}
m_3^2 [ 1+(1-(2 m_3^2)^{-1})^{1/2}]>m_1^2 [ 1+(1-(2 m_1^2)^{-1})^{1/2}]+m_2^2 [ 1+(1-(2 m_2^2)^{-1})^{1/2}]-1/4.
\label{area_increase}
\end{\eq}
Then, the energy emitted in gravitational or other form of radiations is $m_1+m_2-m_3$ and its
efficiency $\ep \equiv (m_1+m_2-m_3)/(m_1+m_2)$ is limited by (\ref{area_increase}).
The highest limit on $\ep$ is $1-3/4\sqrt{2}\approx 0.47 $ which occurs when
$m_1=m_2=2^{-1/2}$, which are the minimum values for positive Hawking temperature $T_H>0$, and $m_3=3/4$. On the other hand, when particles or fields which do not have horizons, impinge on a single black hole, one finds
\begin{\eq}
m_2^2 [ 1+(1-(2 m_2^2)^{-1})^{1/2}]>m_1^2 [ 1+(1-(2 m_1^2)^{-1})^{1/2}].
\label{area_increase_single}
\end{\eq}
{Note that $m_2$ can {\it not} be less than $m_1$. This means that
{\it one can not extract energy from a black hole} and
there is {\it no analogue of the Penrose process} for Kerr or charged black hole in GR \ci{Penr:1969,Chri:1971}. This is basically due to fact that one can not turn-off the parameter
$\om$ arbitrarily, in contrast to the rotation parameter $a$ in a Kerr
black hole or the charge parameter $e$ in a Reissner-Norstr\"{o}m black hole
even though they look similar in the black hole area
formula (\ref{area}) \ci{Hawk:1971}. \footnote{ Actually,
(\ref{area_increase_single}) is exactly the same as that of the Kerr black hole case
with the identification $a=2^{-1}$, whereas
(\ref{area_increase})
as that of Reissner-Norstr\"{o}m black hole
with
$e=2^{-1}$ in \ci{Hawk:1971} {(see \cite{Myun:2009,Cast:2009} for}
some earlier discussions on the similarity).
 }}

These results are {\it quantitatively} (the lower highest limit $\ep =1-2^{-1/2}\approx 0.29$ for two non-rotating initial black holes with the same masses \ci{Hawk:1971}) and {\it qualitatively}
{(no energy extractions via an analogue of the Penrose
process \ci{Penr:1969,Chri:1971})
different from GR black holes which could be tested experimentally in the near future.}\\

8. The ambiguities associated with the choice of asymptotic boundary at $r \ra \infty$ is absent in the charge formula (\ref{Q_Horava}) when we consider $\la=1$ solution (\ref{Park_sol}), where the second and the third (bulk) terms in (\ref{Q_Horava}) vanish. Note that this case covers a wide range of static (vacuum) solutions with higher curvatures, including those in GR \ci{Dola:2018}. However, for $\la \neq 1$ generally \ci{Kiri:2009} which is beyond the GR limit, we still seems to need asymptotic boundary at $r \ra \infty$, even for the asymptotically de Sitter black hole, in order to obtain its ``finite" physical mass. The intimate physical connection between the IR-Lorentz violation and the need of an infinite boundary (of our universe) is still unclear. However, for the asymptotically
de Sitter black hole, the choice of the asymptotic boundary at $r \ra \infty$ might not be
quite nonsensical because
the possible communications between inside and outside of the cosmological horizon from the Lorentz violating effect.\\

9. Based on our proposal of the {off-shell} {\it Diff} invariance which is not spoiled by covariant matter couplings, the manifestly {\it Diff} invariant formulation {\it can be possible} by the change of variables. Its explicit formulation will be interesting because it would be really curious how it differs from the usual covariant higher-curvature gravities. Actually, it reminds us about a covariant formulation of HLD gravity using the Stueckelberg's trick \cite{Germ:2009}. It would be important to see whether their formulation is equivalent to ours or not.\\

\section*{Acknowledgments}
We would like to thank Gokhan Alkac, {Frank Saueressig}, and Sang-Heon Yi for useful correspondences. DOD was supported by the National Natural Science Foundation of China under Grant No. 11875136 and the Major Program of the National Natural Science Foundation of China under Grant No. 11690021. MIP was supported by Basic Science Research Program through the National Research Foundation of Korea (NRF) funded by the Ministry of Education, Science and Technology {(2020R1A2C1010372, 2020R1A6A1A03047877)}.

\appendix

\section{Computational details of (\ref{I0_final})
- (\ref{Omega_i_final}), and explicit form of ${\cal U}^{ij}$}{\label{app1}}

Here, we describe the computational details of (\ref{I0_final})
- (\ref{Omega_i_final}), and the explicit form of ${\cal U}^{ij}$.

First, in order to compute ${\cal I}_0$ in (\ref{I0_final})
from (\ref{I0_general}), without using the dynamical equations of
motion ${\bf E}^{ij}=0$ nor the constraints
${\cal H}\approx 0, {\cal H}^i \approx 0$, we first consider the time derivative $\partial_t ~(\equiv  (~\dot{}~))$
of the potential ${\cal V}$, which appears in the term $N\dot{\cal H}
$ in (\ref{I0_general}),
\begin{align}
\frac{d \mathcal{V}}{d t}=&\frac{\partial \mathcal{V}}{\partial g_{ij}}\partial_t g_{ij}+ \frac{\partial \mathcal{V}}{\partial R^{i}{}_{jkl}}
\partial_t R^{i}{}_{jkl} \no \\
=&\dot{g}_{ij}P^{ilnp}R^{j}{}_{lnp}-P_{i}{}^{jkl}\dot{R}^{i}_{jkl},\label{varV}
\end{align}
where we have used \cite{Wald:1994}
\begin{align}
\left( \frac{\partial {\cal V}}{\partial g^{ij}} \right)_{{R^m}_{nkl}}
=-P_{i}{}^{lnp}R_{jlnp}.\label{suitable}
\end{align}
Expressing the time derivatives
in terms of extrinsic curvature via its definition (\ref{Kij}), we have the relation
\begin{align}
\frac{d \mathcal{V}}{d t}=&\left(2NK_{ij}+\nabla_{i}N_{j}+\nabla_{j}N_{i}\right)P^{ilnp}R^{j}{}_{lnp}
-P_{i}{}^{jkl}\dot{R}^{i}_{jkl} \no \\
=&\left(2NK_{ij}+\nabla_{i}N_{j}+\nabla_{j}N_{i}\right)P^{ilnp}R^{j}{}_{lnp}
-P_{i}{}^{jkl}\left(\nabla_{k}H^{i}{}_{jl}-\nabla_{l}H^{i}{}_{jk}\right),
\end{align}
where we have used a useful relation,
\begin{align}
\dot{R}^{i}_{jkl}=\nabla_{k}H^{i}{}_{jl}-\nabla_{l}H^{i}{}_{jk}
\end{align}
with
\begin{align}
H^{l}_{ij}
\equiv&\nabla_{i}(NK_{j}{}^{l})+\nabla_{j}(NK_{i}{}^{l})-\nabla_{l}(NK_{ij})
+\nabla_{(i}\nabla_{j)}N^{l}-R^{l}{}_{(ij)}{}^{m}N_{m}.\label{dotchris}
\end{align}
With all these identities and (\ref{E})
- (\ref{E_ij_general}),
we can compute (\ref{I0_general}) as
\begin{align}
{\cal I}_{0}=&\nabla_{i}\left\{2N^{2}\left[\nabla_{j}\pi^{ij}+\f{\ka^2}{2} \left( \frac{2\lambda}{\lambda D-1}\left(\pi\nabla_{l}P^{kl}{}_{k}{}^{i}-P^{kl}{}_{k}{}^{i}\nabla_{l}\pi\right)
+2P_{jkl}{}^{i}\nabla^{k}\pi^{jl}
-2\pi^{jl}\nabla^{k}P_{jkl}{}^{i} \right) \right]\right\}
\label{I0new}\\
\equiv &\nabla_{i}{\Omega}^{i},
\end{align}
in terms of the canonical momenta $\pi_{ij}= (2/\ka^2) \sqrt{g} (K_{ij}-\la K g_{ij})$.

As an explicit example, if we consider the potential (\ref{potential_R})
\begin{align}
-\mathcal{V}[g^{ij},R^{i}{}_{jkl}]=&\La+\xi R+\al R^{n}+\beta (R_{ij}R^{ij})^s+\ga ({R^i}_{jkl} {R_i}^{jkl})^r \no \\
=&\La+\xi \delta_{p}^{k}g^{ql}R^{p}{}_{qkl}+\al \,(\delta_{i}^{k}g^{jl}R^{i}{}_{jkl})^{n} \no \\
&+\beta \,(\delta_{i}^{k}R^{i}{}_{jkl}~\delta_{p}^{q}R^{p}{}_{mqn}g^{mj}g^{ln})^s
+\ga ({R^i}_{jkl} {R_i}^{jkl})^r,
\label{genID}
\end{align}
$P_{i}{}^{jkl}$ is given by
\begin{align}
P_{i}{}^{jkl}=
\xi\, \delta_{i}^{[k} g^{l]j}+\al\, n R^{n-1}  \delta_{i}^{[k} g^{l]j}+\beta\, s {\zeta}^{s-1}
(\delta_{i}^{[k}R^{l]j}+g^{j[l}R^{k]}{}_{i})+2\ga r \rho^{r-1}R_{i}{}^{jkl}\label{ptensor},
\end{align}
where we denote ${\zeta} \equiv R_{ij}R^{ij}, \rho \equiv R_{ijkl}R^{ijkl}$.

By plugging (\ref{ptensor}) into (\ref{I0new}), we obtain
\begin{align}
{\cal I}_{0} \equiv & \nabla_{i}{\Omega}^{i}=\nabla_i \left[ \Omega^i_{(0)}+ \tilde{\xi} \Omega^i_{(1)}+\tilde{\al} \Omega^i_{(2)}+\tilde{\be} \Omega^i_{(3)}+\tilde{\ga} \Omega^i_{(4)}\right], \label{I0_explicit}\\
\Omega_{(0)}^{i}= &2N^{2} \nabla_{j}\pi^{ij},
\no \\
\Omega_{(1)}^{i}= &2N^{2}\left[\frac{(\lambda-1)}{(\lambda D-1)}  \nabla^{i}\pi
-\nabla_{j}\pi^{ij}\right],
\no \\
\Omega_{(2)}^{i}=&2nN^{2}\left[\frac{(\lambda-1)}{(\lambda D-1)}\left(R^{n-1}\nabla^{i}\pi
-\pi\nabla^{i}R^{n-1}\right)-\left(R^{n-1}\nabla_{j}\pi^{ij}
-\pi^{ij}\nabla_{j}R^{n-1}\right) \right],
\no \\
\Omega_{(3)}^{i}= & 2sN^{2}\Bigg[\frac{(2\lambda-1)}{(\lambda D-1)}\left(\zeta^{s-1}R^{ij}\nabla_{j}\pi-\pi\nabla_{j}(\zeta^{s-1}R^{ij})\right)
-\frac{\lambda}{(\lambda D-1)}\left(\zeta^{s-1}R\nabla^{i}\pi-\pi\nabla^{i}(\zeta^{s-1}R)\right)\nonumber\\
+&\left(\zeta^{s-1}R^{jk}\nabla^{i}\pi_{jk}-\pi_{jk}\nabla^{i}(\zeta^{s-1}R^{jk})\right)
-\left(\zeta^{s-1}R^{ij}\nabla_{k}\pi_{j}{}^{k}
-\pi_{j}{}^{k}\nabla_{k}(\zeta^{s-1}R^{ij})\right)\nonumber\\
-&\left(\zeta^{s-1}R^{jk}\nabla_{k}\pi^{i}{}_{j}
-\pi^{i}{}_{j}\nabla_{k}(\zeta^{s-1}R^{jk})\right)\Bigg],
\no \\
\Omega^{i}_{(4)}=&2rN^{2}\left[\frac{4\lambda}{\lambda D-1}\left(\pi\nabla_{k}(\rho^{r-1}R^{ik})-\rho^{r-1}R^{ik}\nabla_{k}\pi
\right)+4\rho^{r-1}R_{jkl}{}^{i}\nabla^{k}\pi^{jl}
-4\pi^{jl}\nabla^{k}(\rho^{r-1}R_{jkl}{}^{i})\right],
\no
\end{align}
where $\zeta \equiv R_{ij}R^{ij},~ \rho \equiv R_{ijkl}R^{ijkl}, ~(\tilde{\xi}, \tilde{\alpha}, \tilde{\beta}, \tilde{\gamma})\equiv (\ka^2/2) (\xi, \al, \be, \ga)$. On the other hand, if we consider ${\cal I}_i$ in (\ref{Ik_general}) similarly, one can find that it vanishes identically ${\cal I}_i \equiv 0$, which proves (\ref{Ik_final}), as in GR or general covariant theories.

Similarly, if we consider $\Psi^0$ and $\Psi^i$ in (\ref{Omega0_general}),
(\ref{Omega_i_general}), respectively, one can find that, from (\ref{Theta^0})
and (\ref{Theta_i_general}) as well as (\ref{E})
- (\ref{E_ij_general}),
\begin{align}
\Psi^{0}\equiv&-\xi^{0}\left( N {\cal H}+{N}_{i}{\cal H}^{i}\right)
-\xi^{j}{g}_{ij}{\cal H}^{i} +\Theta^0
\no \\
=&\xi^{0}\mathcal{L}
+\partial_{i}\left[2 \sqrt{g}\, (\xi^{0}N_{j}+\xi_{j}) \left(\f{2}{\ka^2} \right){G}^{ijkl}K_{kl}\right]\label{omega0} \end{align}
and
\begin{align}
\Psi^{i}\equiv & \xi^0 \left[NN^{i}{\cal H}-{\cal H}^{i}(g^{jk}N_{j}N_{k}
-N^{2})+2 N_{j}{\cal H}^{ij} \right] +\xi^j \left( -N_j {\cal H}^{i}+2 g_{jl} {\bf E}^{il}\right)+\Theta^i \no \\
=&\xi^{i}\mathcal{L}+{\bf \Sigma}^{i} (\xi^0)-\partial_{0}\left[2 \sqrt{g}\, (\xi^{0}N_{j}+\xi_{j})   \left(\f{2}{\ka^2} \right){G}^{ijkl}K_{kl}\right]
+\partial_{j}\left[\mathcal{A}^{ij} (\xi^0)+\mathcal{B}^{ij}  (\xi^m)\right],
\label{omegai}
\end{align}
where
\begin{align}
{\bf \Sigma}^{i} (\xi^0)=&
2N^{2}\Bigg[\left(\f{\ka^2}{2} \right)\left( \frac{2\lambda}{\lambda D-1}
\left(P^{li}{}_{lk}\nabla^{k}(\xi^{0}\pi)-\xi^{0}\pi\nabla^{k}P^{li}{}_{lk}\right)
+2\xi^{0}\pi^{jl}\nabla^{k}P_{jkl}{}^{i}-2P_{jkl}{}^{i}\nabla^{k}(\xi^{0}\pi^{jl}) \right)\no \\
+&\pi^{ij}\nabla_{j}\xi^{0}-\xi^{0}\nabla_{j}\pi^{ij}
\Bigg]
\end{align}
and
$\mathcal{A}^{ij}$, $\mathcal{B}^{ij}$ are antisymmetric tensors as
\begin{align}
&\mathcal{A}^{ij} (\xi^0) \equiv 2\sqrt{g}\Big[\left(\f{2}{\ka^2} \right) 2\xi^{0}N_{m}N^{[j}{G}^{i]mkl}K_{kl}
+P^{ijkl}\left(2\xi^{0}N_{l}\nabla_{k}N+N\nabla_{l}(\xi^{0}N_{k})\right)
+4\xi^{0}NN^{l}\nabla^{k}P^{[j}{}_{kl}{}^{i]}\Big], \no \\
&\mathcal{B}^{ij}  (\xi^m) \equiv 2\sqrt{g}\Big[\left(\f{2}{\ka^2} \right) 2\xi_{m}N^{[j}{G}^{i]mkl}K_{kl}+4\xi^{l}\nabla^{k}(N P^{[j}{}_{kl}{}^{i]})+2NP^{[j}{}_{kl}{}^{i]}\nabla^{k}\xi^{l}\Big].
\end{align}
Then, one can write $\Psi^{\mu}\equiv \xi^{\mu} {\cal L}+\Sigma^{\mu} (\xi^0)+\partial_{\nu} {\cal U}^{\mu \nu}$ with $\Sigma^{0}\equiv 0$ and the ``super-potential" ${\cal U}^{\mu \nu}$, which is anti-symmetric ${\cal U}^{\mu \nu}=-{\cal U}^{\mu \nu}$ \cite{Freu:1939,Berg:1949} and given by
\begin{\eq}
{\cal U}^{0i}&=&-{\cal U}^{i0} \equiv  2  \sqrt{g} (\xi^0N_j +\xi_j)  \left(\f{2}{\ka^2} \right) \left(G^{ijkl}  K_{kl} \right),\\
{\cal U}^{ij}&=&-{\cal U}^{ji} \equiv  \mathcal{A}^{ij}(\xi^0)+\mathcal{B}^{ij}(\xi^m),
\end{\eq}
proving (\ref{Omega0_final}) and (\ref{Omega_i_final}).

If we consider the potential (\ref{genID}) with ${P_i}^{jkl}$ tensor (\ref{ptensor}), as an explicit example, one can find ${\bf E}^{ij}$ in (\ref{E_ij_general}) as
\begin{align}
{\bf E}^{ij} \equiv &  E^{ij}_{(0)}+\xi E^{ij}_{(1)}+\alpha E^{ij}_{(2)}+\beta E^{ij}_{(3)}+\gamma E^{ij}_{(4)}, \\
E^{ij}_{(0)}= & \sqrt{g}\, \left( \f{2}{\ka^2} \right) \Big [-N^{i}\nabla_{k}K^{jk}-N^{j}\nabla_{k}K^{ik}+K^{ik}\nabla^{j}N_{k}
+K^{jk}\nabla^{i}N_{k}+N^{k}\nabla_{k}K^{ij}\nonumber\\
+&2NK^{ik}K^{j}{}_{k}-NKK^{ij}+\frac{1}{2}g^{ij}NK^{kl}K_{kl}
-g^{ik}g^{jl}\dot{K}_{kl}\Big ] \no \\
+&\la \sqrt{g}\,\Big [\frac{1}{2}Ng^{ij}K^{2}+N^{j}\nabla^{i}K+N^{i}\nabla^{j}K
-g^{ij}N^{k}\nabla_{k}K-g^{ij}K^{kl}\dot{g}_{kl}+g^{ij}g^{kl}\dot{K}_{kl}\Big ], \no \\
E^{ij}_{(1)}= & \sqrt{g}\, \Big[N \left(-R^{ij}+\frac{1}{2}Rg^{ij}+\frac{\Lambda}{\xi} \, g^{ij}\right)+\left(g^{il}g^{jk}-g^{ij}g^{kl}\right)\nabla_{l}\nabla_{k}N\Big ],\no \\
E^{ij}_{(2)}= &\sqrt{g}\, \, \Big[N \left(-n R^{n-1} R^{ij}+\frac{1}{2} R^{n} g^{ij} \right)+n\left(g^{il}g^{jk}
-g^{ij}g^{kl}\right)\nabla_{l}\nabla_{k}\left(NR^{n-1}\right)\Big], \no \\
E^{ij}_{(3)}= &\sqrt{g}\, \, \Big[ N \left(-2 s \zeta^{s-1} R^{ik}R^{j}{}_{k}+\frac{1}{2} \zeta^s g^{ij}  \right)\nonumber\\
+& s \left(g^{ik} g^{jm} g^{ln}
+g^{jk}g^{mi} g^{nl}-g^{kl} g^{mi} g^{nj}-g^{ij} g^{km} g^{ln}\right)\nabla_{l}\nabla_{k}\left(N \zeta^{s-1} R_{mn}\right)\Big], \no \\
E^{ij}_{(4)}=&\sqrt{g}\left[N \left( -2r\rho^{r-1} R^{iklm}R^{j}{}_{klm}+\frac{1}{2} \rho^{r} g^{ij}\right)
+4r\nabla_{k}\nabla_{l}\left(\rho^{r-1}NR^{iklj}\right)\right] \no,
\end{align}
and
$\Theta^i$ in (\ref{Theta_i_general})
as
\begin{align}
\Theta^{i} \equiv &  \Theta^{i}_{(0)}+\xi \Theta^{i}_{(1)}+\alpha \Theta^{i}_{(2)}
+\beta \Theta^{i}_{(3)}+\gamma \Theta^{i}_{(4)}, \\
\Theta^{i}_{(0)}=&\sqrt{g} \left( \f{2}{\kappa^2} \right)
\left[2N^{l}{G}^{ijkm}K_{km}\delta g_{jl}-N^{i}{G}^{ljmn}K_{mn}\delta g_{jl}
-2{G}^{kjil}K_{kj}\delta N_{l}\right], \no \\
\Theta^{i}_{(1)}=&2\sqrt{g}g^{j[k}g^{l]i}\left[N\nabla_{k}\delta g_{lj}
-(\nabla_{k}N)\delta g_{lj}\right] 
\no ,\\
\Theta^{i}_{(2)}=&4n \sqrt{g}\left[Ng^{i[l}g^{k]j}R^{n-1}\nabla_{k}\delta g_{lj}
-\nabla_{k}(Ng^{i[l}g^{k]j}R^{n-1})\delta g_{lj}\right]
\no ,\\
\Theta^{i}_{(3)}=&4s\sqrt{g}\left[N \zeta^{s-1}g_{[l}{}^{[i}R^{j]}{}_{k]}\nabla^{k}\delta g^{l}{}_{j}-\nabla^{k}(N \zeta^{s-1} g_{[l}{}^{[i}R^{j]}{}_{k]})\delta g^{l}{}_{j}\right], \no \\
\Theta^{i}_{(4)}=&4 r \sqrt{g} \rho^{r-1}R^{jkil}N\nabla_{k}\delta g_{lj}
-4r \sqrt{g} \delta g_{lj}\nabla_{k}(\rho^{r-1}NR^{jikl}). \no
\end{align}

For
${\bf \Sigma}^i$
in (\ref{Sigma_i}), we find
\begin{align}
&{\bf \Sigma}^i (\xi^0) \equiv \Sigma^i_{(0)}+\tilde{\xi} \Sigma^i_{(1)}+\tilde{\alpha} \Sigma^i_{(2)}+\tilde{\beta} \Sigma^i_{(3)}+\tilde{\gamma} \Sigma^i_{(4)}, \\
&\Sigma^{i}_{(0)}=2N^{2}\left( \f{2}{\kappa^2} \right) \left[-\xi^{0}\nabla_{i}\pi^{ij}
+\pi^{i}{}_{j}\nabla^{j}\xi^{0}\right], \no \\
&\Sigma^{i}_{(1)}=2N^{2}\left[-\widehat{\la} \nabla^{i}(\xi^{0}\pi)+\xi^{0}\nabla_{i}\pi^{ij}
+\pi^{i}{}_{j}\nabla^{j}\xi^{0}\right],\no \\
&\Sigma^{i}_{(2)}=2nN^{2}\left[-\widehat{\la}\left(R^{n-1}\nabla^{i}(\xi^{0}\pi)
-\xi^{0}\pi\nabla^{i}R^{n-1}\right)\right],\no \\
&\Sigma^{i}_{(3)}=2sN^{2}\Bigg[\widetilde{\la}\left(\zeta^{s-1}R\nabla^{i}(\xi^{0}\pi)
-\xi^{0}\pi\nabla^{i}(R\zeta^{s-1})\right)\nonumber\\
-& \bar{\la}\left(\zeta^{s-1} R^{ij}\nabla_{j}(\xi^{0}\pi)-\xi^{0}\pi\nabla_{j}(\zeta^{s-1}R^{ij})\right)
+\left(\xi^{0}\pi^{jk}\nabla^{i}(\zeta^{s-1}R_{jk})
-\zeta^{s-1}R_{jk}\nabla^{i}(\zeta^{s-1}R^{ij})\right) \nonumber\\
+&\left(\zeta^{s-1}R_{j}{}^{k}\nabla^{j}(\xi^{0}\pi^{i}{}_{k})
-\xi^{0}\pi^{i}{}_{k}\nabla^{j}(\zeta^{s-1}R_{j}{}^{k})\right)
+\left(\zeta^{s-1}R^{ij}\nabla_{k}(\xi^{0}\pi_{j}{}^{k})
-\xi^{0}\pi_{j}{}^{k}\nabla_{k}(\zeta^{s-1}R^{ij})\right)\Bigg], \no \\
&\Sigma^{i}_{(4)}=2r N^{2}\Bigg[4 \widetilde{\la}\left(\rho^{r-1}R^{ij}\nabla_{j}(\xi^{0}\pi)-\xi^{0}\pi \nabla_{j}(\rho^{r-1}R^{ij})\right) \nonumber\\
&~~~~~~~~~+4\left(\rho^{r-1}R^{i}{}_{jkl}\nabla^{l}(\xi^{0}\pi^{jk})
-\xi^{0}\pi^{jk}\nabla^{l}(\rho^{r-1}R^{i}{}_{jkl})\right)\Bigg], \no
\end{align}
where $\widehat{\la} \equiv (\lambda-1)/(\lambda D-1), \widetilde{\la}\equiv \lambda/(\lambda D-1), \bar{\la} \equiv \widehat{\la}+\widetilde{\la} = (2 \lambda-1)/(\lambda D-1)$,
and $\mathcal{A}^{ij}$, $\mathcal{B}^{ij}$ as
\begin{align}
\mathcal{A}^{ij} (\xi^0) \equiv&  \mathcal{A}^{ij}_{(0)}+\xi \mathcal{A}^{ij}_{(1)}+\al\, \mathcal{A}^{ij}_{(2)}
+\beta\, \mathcal{A}^{ij}_{(3)}+\ga\, \mathcal{A}^{ij}_{(4)}, \\
\mathcal{A}^{ij}_{(0)}=&2\sqrt{g} \left(\frac{2}{\kappa^2}\right)\left[ 2\xi^{0}N_{m}N^{[j}G^{i]mkl}K_{kl}\right], \no\\
\mathcal{A}^{ij}_{(1)}=&2\sqrt{g}\left[g^{i[k}g^{l]j}\left(2\xi^{0}N_{l}\nabla_{k}N
+N\nabla_{l}(\xi^{0}N_{k})\right)\right],\no \\
\mathcal{A}^{ij}_{(2)}=&2n\sqrt{g}\left[R^{n-1}g^{i[k}g^{l]j}\left(2\xi^{0}N_{l}\nabla_{k}N
+N\nabla_{l}(\xi^{0}N_{k})\right)+4\xi^{0}NN_{l}g^{i[l}g^{k]j}\nabla_{k}R^{n-1}\right], \no\\
\mathcal{A}^{ij}_{(3)}=&2s\sqrt{g}\left[2s\zeta^{s-1}g^{[j}_{[l}R^{i]}_{k]}
\left(2\xi^{0}N^{l}\nabla^{k}N+N\nabla^{l}(\xi^{0}N^{k})\right)
+8\xi^{0}NN^{l}g^{[i}_{[l}\nabla^{|k|}(\zeta^{s-1}R^{j]}_{k]})\right], \no \\
\mathcal{A}^{ij}_{(4)}=&
2r\sqrt{g}\Big[8\xi^{0}NN^{l}\nabla^{k}\left(\rho^{r-1}R^{[j}{}_{kl}{}^{i]}\right)
+2\rho^{r-1}R^{ijkl} \left(2\xi^{0}N_{l}\nabla_{k}N+N\nabla_{l}(\xi^{0}N_{k}) \right)
\Big], \no
\end{align}
and
\begin{align}
\mathcal{B}^{ij} (\xi^m) \equiv&\mathcal{B}^{ij}_{(0)}+\xi \mathcal{B}^{ij}_{(1)}+\al\, \mathcal{B}^{ij}_{(2)}
+\beta\, \mathcal{B}^{ij}_{(3)}+\ga\, \mathcal{B}^{ij}_{(4)},\label{Btensor}\\
\mathcal{B}^{ij}_{(0)}=&2\sqrt{g} \left(\frac{2}{\kappa^2}\right) \left[2\xi_{m}N^{[j}G^{i]mkl}K_{kl}\right], \no \\
\mathcal{B}^{ij}_{(1)}=&2\sqrt{g}\left[2g^{l[i}g^{j]}{}_{k}(N\nabla^{k}\xi_{l}
-2\xi_{l}\nabla^{k}N)\right], \no\\
\mathcal{B}^{ij}_{(2)}=&2n \sqrt{g}\left[2g^{l[i}g^{j]}{}_{k}R^{n-1}N\nabla^{k}\xi_{l}
+4\xi_{l}\nabla^{k}(g^{l[j}g^{i]}{}_{k}R^{n-1}N) \right], \no\\
\mathcal{B}^{ij}_{(3)}=&2s \sqrt{g}\left[4\zeta^{s-1}g^{[j[k}R^{i]l]}N\nabla_{k}\xi_{l}
+8\xi_{l}\nabla_{k}(\zeta^{s-1}g^{[j[k}R^{i]l]}N)\right], \no\\
\mathcal{B}^{ij}_{(4)}=&2r \sqrt{g}\left[4\rho^{r-1}R^{k[ij]l}N\nabla_{k}\xi_{l}
+8\xi_{l}\nabla_{k}(\rho^{r-1}R^{k[ji]l}N)\right]. \no
\end{align}

\section{A Summary of the Complete Set of Constraints Obtained in \cite{Deni:2020} $(\la \neq 1/D)$}{\label{app2}}

Here, we summarize the complete set of constraints for the Hamiltonian formalism, obtained in \cite{Deni:2020} $(\la \neq 1/D)$ (see \cite{Deni:2020} for the detailed computations and the case of $\la = 1/D$). In \cite{Deni:2020}, we have considered the HLD action (\ref{action1}) with a potential ${\cal V}(R)$, which is an arbitrary function of curvature scalar $R$, or more explicitly,
$
-\mathcal{V}= \La+ \xi R+\al R^n,
$
for the computational simplicity. Then, from the {\it primary} constraints\
\begin{\eq}
\pi_N \equiv \de S/\de \dot{N} \approx 0,~\pi^i \equiv \de S/\de \dot{N_i} \approx 0
\label{primary_cons}
\end{\eq}
and their preservation $\dot{\Phi}^{\mu} =\{\Phi^{\mu}, H_C \}\approx 0$ [$\Phi^\mu \equiv (\pi_N , \pi^i)$] with
the canonical Hamiltonian (up to boundary terms)
\begin{\eq}
\label{Hc}
 H_C&=& \int_{\Si_t} d^{D}x  \left\{ N {\cal H}+N_i {\cal H}^i\right\},
\end{\eq}
one obtain
the {\it secondary} constraints
\begin{\eq}
\label{secondary_const}
{\cal H} \approx 0,~~
{\cal H}^i \approx 0,
\end{\eq}
which are the Hamiltonian and momentum constraints, for ${\cal H}$ and ${\cal H}^i$ in (\ref{E}) and (\ref{E1}), respectively. Here, the weak equality `$\approx$' means that the constraint equations are used after calculating the Poisson brackets. Up to now, the constraint analysis looks parallel with GR but now with the modified expression of ${\cal H}$ as (\ref{E}). However, we note that ${\cal H}^i$ has still the same expression as in GR in terms of canonical momenta $\pi^{ij}=(2/\kappa^2)\sqrt{g} (K^{ij}-\lambda K g^{ij})$, though different from in terms of $g^{ij}$ and $\dot{g}^{ij}$, or $K^{ij}$. This fact indicates the same role of ${\cal H}^i$ as in GR in Hamiltonian formalism, due to the fundamental role of $\pi^{ij}$.

The fundamentally different constraint analysis in HLD gravity than in GR starts from the different constraint algebra
\begin{\eq}
 \{ {\cal H} (x),{\cal H} (y) \}&=&C^i (x) \nabla^x_i \de^D (x-y)-C^i (y) \nabla^y_i \de^D (x-y), \label{HtHt}\\
 \{ {\cal H} (x),{\cal H}_i (y) \}&=&-{\cal H} (y) \nabla^y_i
 \de^D (x-y), \label{HtHi}\\
 \{ {\cal H}_i (x),{\cal H}_j (y) \}&=&{\cal H}_i  (y) \nabla^x_j \de^D (x-y)+{\cal H}_j  (x) \nabla^x_i \de^D (x-y), \label{HiHj}
\end{\eq}
where
$C^i$ is the same quantity in (\ref{C}) but with $\beta=\ga=0$, as in ${\cal H}$ (\ref{E}).
Then, the preservation of the secondary constraints with the {\it extended} Hamiltonian
$
 H_E
 =
 H_C + \int_{\Si_t} d^{D}x~(   u_{\mu} \Phi^{\mu} ),
$
 with the Lagrange multipliers $u_{\mu}$ due to the arbitrariness from the primary constraints,
\begin{\eq}
\dot{{\cal H}}&\equiv &\{{\cal H} , H_E \} \no \\
&=&\f{1}{N} \nabla_i (N^2 C^i) +\nabla_i (N^i {\cal H}) \approx 0
, \label{H_t_dot} \\
\dot{{\cal H}_i}&\equiv &\{{\cal H}_i , H_E \} \no \\
&=&{\cal H} \nabla_i N +\nabla_j (N^j {\cal H}_i)+{\cal H}_j \nabla_i N^j \approx 0
\label{H_i_dot}
\end{\eq}
produce the {\it tertiary} constraint,
\begin{\eq}
\Omega \equiv \nabla_i (N^2 C^i) +N \nabla_i (N^i {\cal H}^t)\approx 0,
\label{Omega_const}
\end{\eq}
excluding the trivial case of $N=0$ for all space-time.

One more step of preserving the tertiary constraint
gives
\begin{\eq}
\dot{{\Omega}}&\equiv& \{{\Omega}, H_E\} \no \\
&\approx & \{{\Omega}, H_C\}+2 {C}^i N^2 \nabla_i \left( \f{u_t}{N} \right) \approx 0.
\label{Om_dot}
\end{\eq}
Now, the remaining further analysis depends on whether
${C}^i \approx 0$ or ${C}^i \not\approx 0$.

{\bf A}. Case ${C}^i \approx 0$: In this case, the multiplier $u_t$ is not determined in (\ref{Om_dot}) but we need one more step with a further constraint $\Xi \equiv \{{\Om} , H_C \}$ (see \cite{Deni:2020} for the explicit expression and the more details). Then, the full set of constraints is given by $\chi_A \equiv ( \pi_N, {\cal H}, {\Om}, {\Xi} )\approx 0, \Ga_B \equiv (\pi^i, {\cal H}_i ) \approx 0$. Here,  the constraints
$\chi_A  \approx 0$ are the
{\it second-class} constraints with the constraint algebra,
\begin{\eq}
&&\{  \pi_N (x), {\cal H} (y) \} =0 , \no \\
&&\{   \pi_N (x), {\Om} (y) \} \approx -2 \nabla^y_i \left( N {C}^i(y) \de^D(x-y)\right)\approx 0, \no \\
&&\{  \pi_N (x), \tilde{\Xi} (y) \} =\Delta (x-y), \no \\
&&\{ {\cal H} (x), {\cal H} (y) \} =C^i (x) \nabla^x_i \de^D (x-y)-C^i (y) \nabla^y_i \de^D (x-y)\approx 0 , \no \\
&&\{ {\cal H} (x),  \tilde{\Omega} (y) \} \approx \{  \pi_N (x), {\Xi}^i (y) \} ,~ etc,
\label{const_Poisson_2_local_text}
\end{\eq}
whose determinant $det(\{\chi_A, \chi_B \})$ is non-vanishing generally.

On the other hand, the constraints
$\Ga_A \equiv (\pi^i, {\cal H}_i ) \approx 0$
are the {\it first-class} constraints with the vanishing 
determinant
$det (\{\Ga_A, \Ga_B \} )=0$.
Then, the number of dynamical degrees of freedom in the ``configuration" space is given by
\begin{\eq}
s&=&\f{1}{2} \left( P-2 N_1-N_2\right) \no \\
&=&\f{1}{2} (D+1)(D-2),
\end{\eq}
where $P=(D+1)(D+2)$ is the number of canonical variables in ``phase" space $(N, \pi_N, N_i, \pi^i,g_{ij}, \pi_{ij})$, $N_1=2D$ is the number of the first-class constraints $(\pi^i, {\cal H}_i)\approx 0$, and $N_2=``4"$ is the number of the second-class constraints
$(\pi_N, {\cal H}, {\Om}, {\Xi} )\approx 0$. It is remarkable that the {\it 2 first-class} constraints
$(\pi_N, {\cal H})\approx 0$ in GR
transform into the {\it 4 second-class} constraints
$( \pi_N, {\cal H}, {\Om}, {\Xi} )\approx 0$ in the
Case {A} of HLD gravity,
maintaining the same dynamical degrees of freedom $s$.

{\bf B}. Case ${C}^i \not\approx 0$: In this case, the multiplier $u_t$ is now determined in (\ref{Om_dot}) and there is no further constraint. Then, in contrast to Case {A}, there are the second-class constraints ${\chi}_A \equiv ( \pi_N, {\cal H}, \Om )\approx 0$, whose determinant
$det(\{{\chi}_A, {\chi}_B \})$ is generally non-vanishing,
whereas the first-class constraints $\Ga_A \equiv (\pi^i, {\cal H}_i )$ are the same as in the Case {A}. So, the number of dynamical degrees of freedom is
\begin{\eq}
s&=&\f{1}{2} \left[ (D+1)(D+2)-2 \times 2 D -``3"\right] \no \\
&=&\f{1}{2} (D+1)(D-2)  +\f{1}{2},
\end{\eq}
with $N_1=2 D$ and $N_2=``3"$, which shows {\it one-extra} degree of freedom in phase space, in addition to the usual $(D+1)(D-2)$ graviton (transverse traceless) modes in GR or the Case {A} of HLD gravity in arbitrary $(D+1)$ dimensions. This result supports the previous case-by-case results \ci{Bell:2011,Deve:2018}
but in a more generic set-up.

\end{document}